# Kondo Effect in Micron Size Device Fabricated From Flakes of Mn Doped $Bi_2Se_3$ Topological Insulator

Vishal K. Maurya[*,2], Jeetendra K. Tiwari[1], S. Ghosh[1] and S. Patnaik[1]

[1] Jawaharlal Nehru University, New Delhi, India, 110067
[2]CSIR, National Physical Laboratory, K.S. Krishnan Marg, New Delhi, 110012 *mauryavishal1@gmail.com

**Abstract.** Single crystals of $Mn_{0.03}Bi_{1.97}Se_3$ were synthesized by modified Bridgman technique and phase purity was confirmed via XRD analysis. EDAX analysis has verified the stoichiometric ratio of elements in the sample. Sample flakes were transferred to the $SiO_2$/Si n-type substrate by mechanical exfoliation technique. Four probe gold contacts were etched with the help of e-beam lithography by masking and lift off process. Resistivity measurement was performed in four probe configurations in 2-300 K temperature range. We report evidence for Kondo effect in $Mn_{0.03}Bi_{1.97}Se_3$ micro-flakes with $T_{min}$ of 14.4 K.

## 1 Introduction

Nuances of topological quantum states in surfaces of 3D topological insulators (TI) such as $Bi_2Se_3$ $Bi_2Te_3$ and $Sb_2Te_3$ have attracted considerable attention in the recent past [1, 2]. These properties of Topological Insulators (TI) make them interesting for studying quantum phenomena that can lead to detection of Majorana zero modes, applications in quantum computations and spintronics, Josephson junction, heterostructure effect [3,4,5]. There are various methods to detect surface states such as Spin Resolved Photo Emission Spectroscopy (SRPES), Magnetic Probe Scanning Tunneling Microscope (STM), and thickness-based charge transport measurements [6, 7]. These TI systems have recently been observed for many other exotic quantum phenomena as Shubnikov de Haas (SDH) oscillations, spin charge detection and Weak anti Localization (WAL) effect [8-10]. Transport measurements in TIs are challenging because of the semi-metallic nature of the bulk (non-zero bulk resistance). The resultant quantum coherence length vs. sample size limitations demands fabrication of micron size thin devices for surface transport measurements at low temperatures. Here in this paper we report the successful growth of $Mn_{0.03}Bi_{1.97}Se_3$ single crystals and fabrication of micron size four probe devices suitable for electronic transport measurements at low temperatures.

## 2  Sample preparation

The growth of single crystals of $Mn_{0.03}Bi_{1.97}Se_3$ was carried out by melting stoichiometric amounts of high purity elements of Bi (99.99%), Se (99.99%) and Mn(99.99%) at 750 °C for 3 days in sealed evacuated quartz glass tube. Tube was shaken from time to time to get homogeneous sample. Sample was cooled slowly to 550 °C over 24 h followed by annealing at same temperature for 72 h. Quenching was done in cold water.

In Fig. 1 XRD data is shown for sample crystals flakes, clearly showing only (0, 0, l) peaks which is an indication of good quality and well oriented crystals along c axis. XRD pattern is matched by ICDD data card no. 33-0214 using Philips X'pert High Score software. The lattice parameters of sample are found to be a=b=4.13 Å and c = 28.63 Å with rhombohedral structure and R-3m space group. In the inset (a) of figure 1, images of crystal flakes on a graph paper are shown. In the inset (b) of figure, 1, SEM image of broken edge surface layers is displayed at 10 μm resolution scale. Layered morphology of surface is evident.

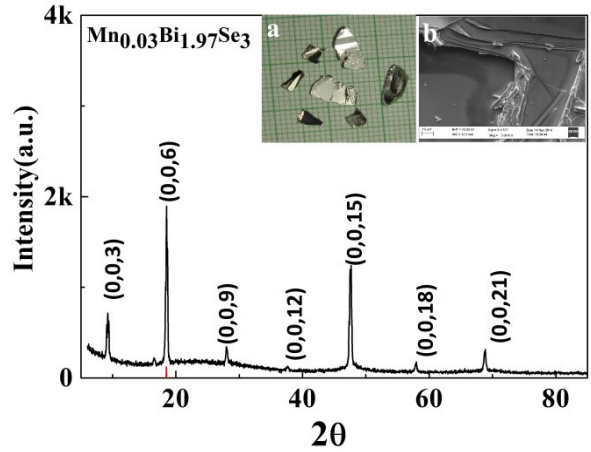

**Fig. 1.** The XRD pattern for $Mn_{0.03}Bi_{1.97}Se_3$ shows only (0, 0, *l*) peaks. In the inset (a) as grown single crystals (cm size) on a graph sheet are shown. One small green box represents 1 $mm^2$ area. In inset (b) SEM image of the single crystal is shown.

## 3  Device Fabrication

Electron beam lithography is a convenient technique to make micron-nano scale devices. 5×5 mm pieces of $SiO_2$/Si ($n^{++}$) substrate were cleaned in deionized water followed by trichloro ethylene, acetone and finally in ethanol for ten minutes each by sonication. $Mn_{0.03}Bi_{1.97}Se_3$ sample flakes were exfoliated and transferred to $Si/SiO_2$ substrate. A strip of about 8 cm in length and 2 cm in width of scotch tape is used for the mechanical exfoliation of sample flakes. Multiple folding of the scotch tape ensured the desired thickness of the flakes. Flakes were first analysed under an optical microscope and the followed by SEM for thickness, size and smoothness of surface. Pictures of flakes under an optical microscope are shown in Fig. 2 (a) and SEM images are shown in Fig. 2 (b).

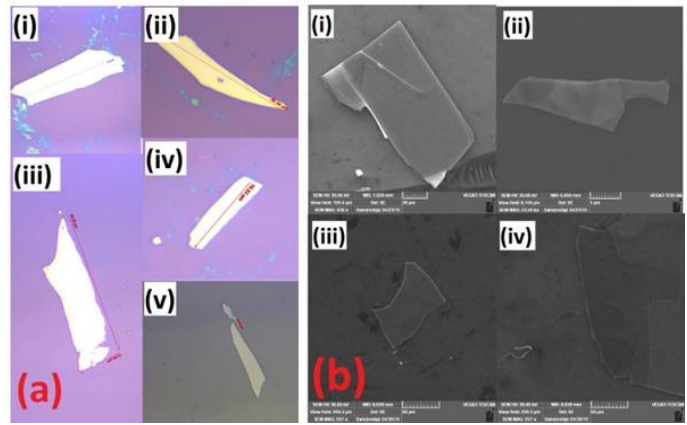

**Fig. 2.** (a) Flakes of $Mn_{0.03}Bi_{1.97}Se_3$ transferred to the $SiO_2$/Si substrate by mechanical exfoliation with the help of *Scotch* tape. (b) SEM images of sample flakes.

After successful transfer of the flakes onto the substrate, a homogenous layer of Poly Methyl Methacrylate (PMMA) dissolved in anisole was coated on the exfoliated flake using a spin coater. After spin coating sample was baked at 180 $^0$C for 90 sec to form a homogeneous layer. After spin coating of PMMA layer, substrate was put into the SEM chamber for the etching process. A beam of electron is incident on the PMMA layered substrates. The electron beam changed the solubility of PMMA only at the incident area and rest area remained intact. After etching desired pattern, substrate was put in the developing solution (mixture of methyl iso-butyl ketone and iso-propyl alcohol in 1:3 ratios) for 90 seconds. Finally substrate was rinsed into iso-propyl alcohol. Thus, etching and development was completed. After developing the PMMA contact pattern, we used thermal evaporation technique to deposit gold (Au) on the layer. Sample was put into an evacuated chamber and metal was evaporated in a controlled manner so that a few nano-meter film of Au could be deposited. Substrate with Au contacts was put into acetone to remove all the remaining PMMA content. In Fig. 3 (a) final device with four probe contacts is shown in Figure 3 (b) a magnified image of same device is shown. Each contact width was ~7 μm in width and the separation between two consecutive contacts was ~ 7 μm. Sample flake dimensions, determined by SEM, were 50×30 μm.

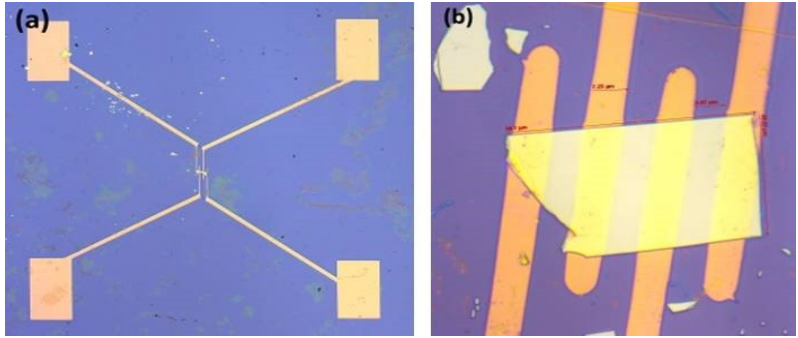

**Fig. 3.** (a) Image of the fabricated device with four probe contacts. Contacts pads can be seen in golden color. (b) A 50 times magnified image of sample with four probes in series is depicted.

## 4   Transport Measurement

Resistivity measurement was done in zero magnetic field in 2-300 K temperature range. Sheet resistance in micro-flakes is almost linear in high temperature region but it shows a nominal upturn at low temperature below 14 K. We have examined the behavior of temperature dependency of resistance in two different temperature regions described by the two following equations

$$R_s(T) = a_0 + a_1 T^2 + a_2 \ln T \; ; \; (2 \text{ - } 35\text{K}) \tag{1}$$

$$R_s(T) = a_0 + a_3 T + a_1 T^2 \; ; \; (35 \text{ - } 300\text{K}) \tag{2}$$

The low temperature data fitting is shown by the red solid line and high temperature data fitting is shown by the green solid line in the Fig. 4

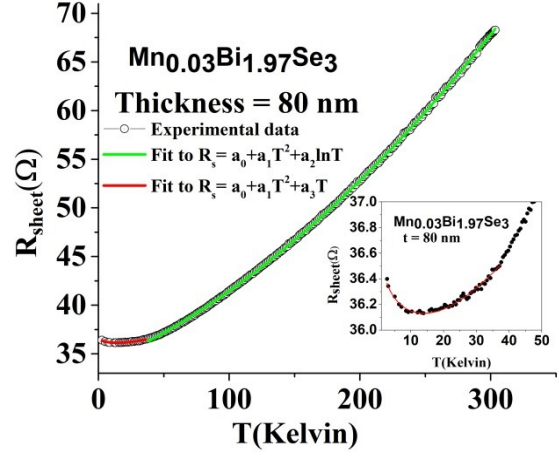

**Fig. 4.** Sheet resistivity data is fitted in the 2-35K and 35-300K temperature ranges for Mn$_{0.03}$Bi$_2$Se$_3$ micro device. In the inset magnified 0-35 K resistivity data is shown.

The $a_0$ = 36.55 Ω, represents the temperature independent term which correspond to the residual sheet resistance. For the low temperature range data, the coefficient of $T^2$ term ($a_1$ = 4.6×10$^{-4}$ Ω/K$^2$), known as Fermi liquid term, represents the resistivity contribution due to electron-electron scattering at low temperatures. The coefficient for the term $a_3$ which represents the linear metallic behavior of resistivity has an appreciable magnitude of 5.8×10$^{-2}$ Ω/K. The logarithmic term represents the Kondo effect, is due to Mn doping which acts as magnetic scattering in this system. It is also observed in semi metals, quantum dots and heavy Fermionic systems [11, 12].

Kondo Temperature is estimated by the following equation.

$$T_{min} = \left(\frac{|a_2|}{2a_1}\right)^{1/2} \quad (3)$$

We estimated the value of $T_{min}$ = 14.4 K. where $a_2$ = -0.19 Ω and $a_1$ = 4.6×10$^{-4}$ Ω/K$^2$ from the fitting of data by equation 1 and 2.

## 5   Conclusion

In conclusion, we have successfully synthesized single crystals of Mn$_{0.03}$Bi$_{1.97}$Se$_3$. XRD, EDX and SEM analyses were performed to check the phase purity of the sample, stoichiometric ratio, the surface morphology, respectively. Observation of (0 0 *l*) peaks indicates sample of being highly layered and oriented along c axis. A micron sized device is fabricated for four probe resistivity measurements in the temperature range 2-300 K under zero magnetic field. Sheet resistance data produced a good fit for two different temperature ranges: 2-35 and 35-300 K. Kondo effect is observed in $Mn_{0.03}Bi_{1.97}Se_3$ flake with $T_{min}$ at 14.4 K.

# 6 Acknowledgement

VKM, and JKT Thank UGC for Scholarship. SP thanks SERB (project ID EMR/2016/003998/PHY) and DST (Project ID DST/NM/TUE/QM10/2109(G)/6).